\documentclass[namedreferences]{solarphysics}
\usepackage[optionalrh,solaromanenum]{spr-sola-addons} 
\usepackage{graphicx}                    
\usepackage{color}                       
\usepackage{url}                         

\usepackage[pdfborder={0 0 0 },urlcolor=blue]{hyperref}
\ifx \adsurl  \undefined \def \adsurl#1{\href{http://adsabs.harvard.edu/abs/#1}{\textsf{#1}}}\fi

\begin{document}

\begin{article}

\begin{opening}

\title{Predictions of the Maximum Amplitude, Time of Occurrence, and Total Length of Solar Cycle 24}

\author{L.C.~\surname{Uzal}$^{1}$\sep
        R.D.~\surname{Piacentini}$^{2,3}$\sep
        P.F.~\surname{Verdes}$^{1}$      
       }


\runningauthor{L.C.~\surname{Uzal} {\it et.\ al.}}
\runningtitle{Prediction of Solar Cycle 24}

\institute{$^{1}$ French-Argentine International Centre for\\Information and Systems Sciences (CIFASIS)\\
UPCAM (France) / UNR -- CONICET (Argentina) \\
Boulevard 27 de Febrero 210 Bis, S2000EZP Rosario, Argentina
      email: \href{mailto:uzal@cifasis-conicet.gov.ar}{uzal@cifasis-conicet.gov.ar} \\
      email: \href{mailto:verdes@cifasis-conicet.gov.ar}{verdes@cifasis-conicet.gov.ar} \\
             $^{2}$ Instituto de Fisica Rosario (IFIR)\\
UNR -- CONICET (Argentina) \\
Boulevard 27 de Febrero 210 Bis, S2000EZP Rosario, Argentina
       email: \href{mailto:piacentini@ifir-conicet.gov.ar}{piacentini@ifir-conicet.gov.ar} \\
$^{3}$ Facultad de Ciencias Exactas, Ingenieria y Agrimensura \\
Universidad Nacional de Rosario \\
Av.\ Pellegrini 250, S2000BTP Rosario, Argentina
             }

\begin{abstract}
In this work we predict the maximum amplitude, its time of occurrence, and the total length of Solar Cycle 24 by linear regression to the curvature (second derivative) at the preceding minimum of a smoothed version of the sunspots time series.  We characterise the predictive power of the proposed methodology in a causal manner by an incremental incorporation of past solar cycles to the available data base.  In regressing maximum cycle intensity to curvature at the leading minimum we obtain a correlation coefficient R $\approx$ 0.91 and for the upcoming Cycle 24 a forecast of 78 (90\% confidence interval: 56\,--\,106).  Ascent time also appears to be highly correlated to the second derivative at the starting minimum (R $\approx$ -0.77), predicting maximum solar activity for October 2013 (90\% confidence interval: January 2013 to September 2014).  Solar Cycle 24 should come to an end by February 2020 (90\% confidence interval: January 2019 to July 2021), although in this case correlational evidence is weaker (R $\approx$ -0.56).
\end{abstract}


\end{opening}


\section{Introduction}
\label{s:introduction} 

The main features of solar magnetic activity, as measured by the solar (or Wolf) sunspots number (SSN), are successfully explained by dynamo theory \cite{Ossendrijver2003}.  A short cycle was first discovered by Schwabe and later reported by Wolf, who estimated its period to be approximately 11 years.  This cycle is not periodic, with variations in amplitude and length, and also phases of inactivity such as the Maunder minimum.  Longer-scale structures known as Gleissberg cycles have also been identified, but their analysis is complicated by the limited time-span of observations.  The existence of multiple-scale dynamics, or non-stationarity, adds a layer of complexity to the important problem of solar-activity forecasting.  This problem is technologically relevant as it affects the estimation of orbital drag and other space-weather effects such as the rate of solar flares, coronal mass ejections, and cosmic rays.

Different methods have been proposed in the literature to predict the maximum amplitude of Solar Cycle 24 (for a good review see \citeauthor{Pesnell2008}, 2008).  Following \inlinecite{Pesnell2008}, existing techniques can be broadly classified as climatological, precursor, dynamo model, spectral, and neural network:
\begin{itemize}
\item Climatological forecasts assume that amplitude variations around the mean are random and therefore predict a forthcoming cycle of mean intensity, where the average can be computed over the full historical record or a more recent past \cite{Wang2002,Horstman2005,Kim2006,Lantos2006,Clilverd2006}.  
\item Precursor methods have recourse to leading indicators of solar activity.  Two important examples of this type of signals are the intensity of solar polar magnetic fields and geomagnetic activity at the declining phase or cycle minimum \cite{Schatten2005,Svalgaard2005,Hathaway2006,Jain2006,Hamid2006,Kane2007-2}.  
\item Dynamo models, on the other hand, attempt to explain solar dynamics from first principles by integrating conservation equations.  Examples of this type of approach include \inlinecite{Dikpati2006} and \inlinecite{Choudhuri2007}. 
\item Spectral methods search for regularities or predictable patterns in the spectral content of the SSN time series.  Autoregressive forecasts are also classified as spectral.  Representative examples are given by \inlinecite{Kane1999}, \inlinecite{Duhau2003}, \inlinecite{deMeyer2003}, \inlinecite{Clilverd2006}, and \inlinecite{Hiremath2008}.
\item Finally, neural networks attempt to capture nonlinear relationships between a set of input features and future solar activity.  Examples in this category are given by \inlinecite{Calvo1995}, \inlinecite{Verdes2004}, and \inlinecite{Maris2006}.
\end{itemize}

The proposed method is statistical in nature and falls into the climatological class.  More precisely, we apply and extend a previously developed methodology for maximum solar-activity prediction \cite{Verdes2000}.  Based on the curvature of the SSN time series at a solar-cycle minimum, we forecast the main features of the following cycle, namely maximum amplitude, time to maximum, and total length.  In \inlinecite{Verdes2000} we proposed a restricted version of this approach only for the prediction of maximum amplitude and applied it to the case of Solar Cycle 23.  The contributions of the present work are: i) the application of the previously proposed technique to predict the maximum intensity of Solar Cycle 24, and ii) its extension to forecast other cycle features such as ascent time and total length.  The rationale behind the proposed method can be ultimately summarised as follows: fast-rising cycles will reach higher levels than slower-rising ones, with shorter ascent times and total cycle lengths.  This is known as the Waldmeier effect \cite{Waldmeier1955} and has been further documented by \inlinecite{Hathaway1994}, \inlinecite{Hathaway2002}, and \inlinecite{Dikpati2008}.
As a proxy for rising speed we use the curvature (second derivative) at the minima of the SSN time series, and effectively find that it determines the main characteristics of the unfolding cycle.  We report details on these studies below.

\section{Data Preparation and Predictor Signal Construction}
\label{s:data} 

We used the monthly averages of sunspot counts published by the Solar Influences Data Analysis Center (SIDC), which is the Solar Physics research department of the Royal Observatory of Belgium (try \href{http://sidc.oma.be/DATA/monthssn.dat}{http://sidc.oma.be/DATA/ monthssn.dat}). 

\begin{figure} 
\centerline{\includegraphics[width=\textwidth]{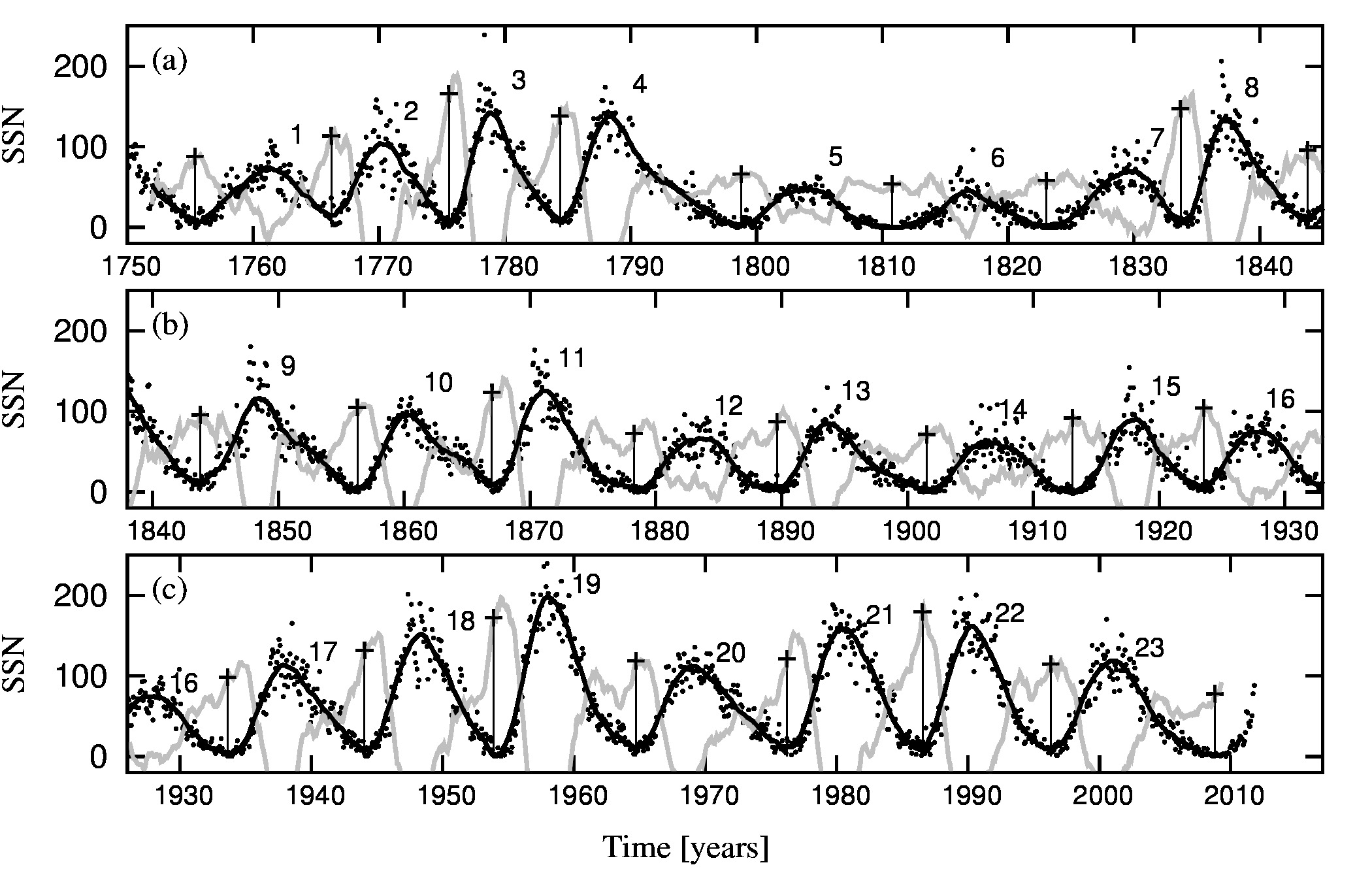}}
\caption{Construction of the signal used to predict the main features of solar cycles.  Raw monthly data are indicated with scattered dots.  The series $d_0(t)$ and $d_2(t)$ obtained from a Savitzky--Golay filter are depicted with solid black and gray lines, respectively.  Vertical lines highlight the values of $d_2(t)$ at the minima of $d_0(t)$, which is the key signal used in this work to forecast shape parameters of solar cycles. 
A second axis is used for $d_2(t)$, which has been conveniently scaled in order to visually maximise the correspondence between the proposed predictor values and subsequent maxima of $d_0(t)$, which is the essence of our method.  Notice also that for visual clarity we do not show negative $d_2(t)$ values as they are unused by the proposed forecasting approach. Cycles are annotated by their number.
}
\label{fig:data}
\end{figure}

In \inlinecite{Verdes2000} we smoothed the raw data with a nonlinear technique known as local projective noise reduction \cite{Grassberger1993}.  Since the proposed methodology is independent of the specific smoothing procedure, here we followed a simpler approach and used a Savitzky--Golay filter \cite{Savitzky1964} fitting parabolas in windows of 61 months (30 months to each side of the point being smoothed).  The Savitzky--Golay filter allows us to estimate a smooth curve interpolating the sunspot data and its second derivative everywhere.  In the following we call these time series $d_0(t)$ and $d_2(t)$, respectively.  Finally, we construct our predictor signal by searching for the 24 minima of $d_0(t)$ and evaluating $d_2(t)$ at these points.  This process is illustrated in Figure \ref{fig:data}.

\section{Predicting the Main Shape Parameters of Solar Cycles}
\label{s:prediction} 

\subsection{Maximum Amplitude}
\label{ss:maximumAmplitudePrediction} 

To predict the maximum amplitude of a solar cycle, we chose as a target variable the yearly averages of sunspot numbers also published by the Solar Influences Data Analysis Center (SIDC, 
\href{http://sidc.oma.be/DATA/yearssn.dat}{http://sidc.oma.be/DATA/yearssn.dat}).  More precisely, we focus on the last 23 maxima of this time series.

\begin{figure} 
\centerline{\includegraphics[width=0.65\textwidth]{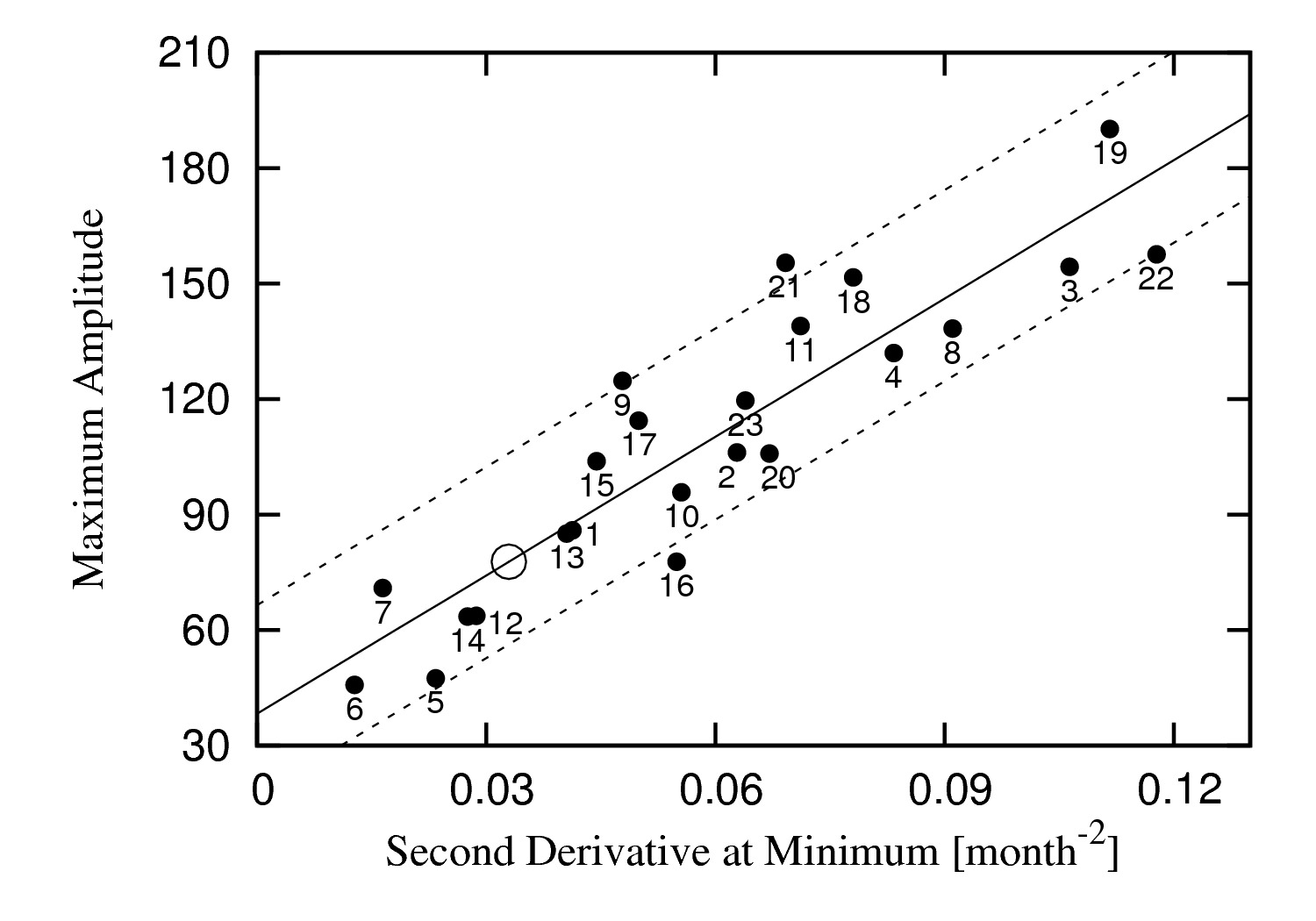}}
\caption{Maximum amplitude of solar cycles against the value of the second derivative at the preceding minimum.  Full and dashed lines indicate a robust linear fit and its 90\% confidence interval, respectively.  Dots are annotated by the corresponding solar cycle number.  The open circle indicates the prediction for Solar Cycle 24.}
\label{fig:linearFitMaximumAmplitude}
\end{figure}

In Figure \ref{fig:linearFitMaximumAmplitude} we show a scatter plot of solar-cycle maxima as a function of the second derivative at the leading minima.  With solid and dashed lines, we plot a  fitted linear model and the 5 and 95 quantiles of its residues, respectively.  
We employed a robust fitting procedure that minimises the median absolute deviation (MAD) \cite{Huber1981}.
We obtain a correlation coefficient R $\approx$ 0.91, and for the upcoming Cycle 24 a forecast of 78 (shown with an open circle) and a 90\% confidence interval of 56\,--\,106 (dashed lines).  The standard error (sample standard deviation of residuals) is 16.

\begin{figure} 
\centerline{\includegraphics[width=0.75\textwidth]{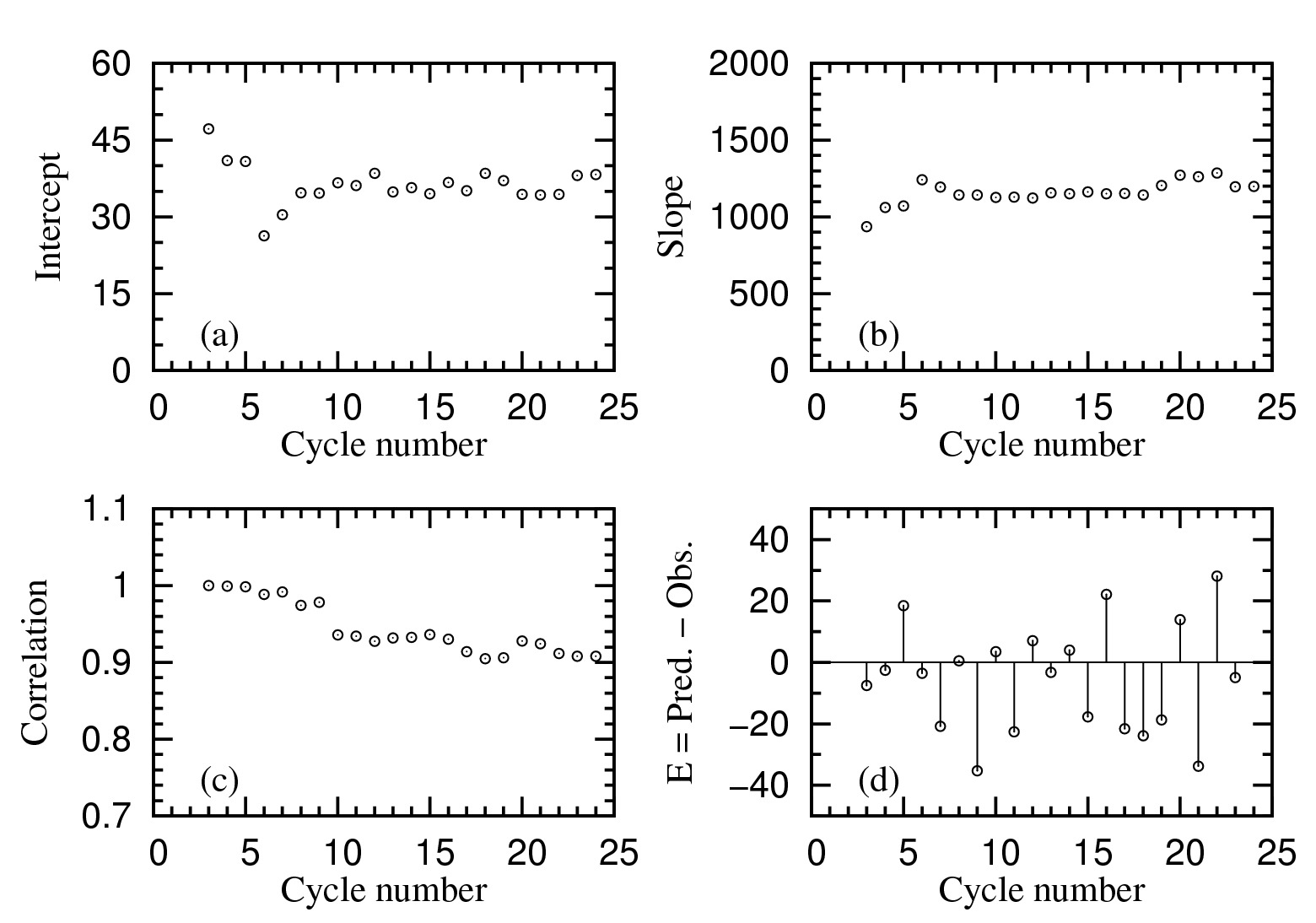}}
\caption{Characterisation of the proposed forecasting methodology.  In panel (a) we show the evolution of the linear regression-intercept as new cycles are added to the data base (more precisely, as a function of the predicted solar cycle number). Panel (b) profiles the slope behaviour. In (c) we plot the evolution of the correlation coefficient between maximum amplitude and curvature at the preceding minimum.  The linear relationship is found to be stable over time, with coefficients roughly unchanged from Cycle 10 onwards.  Finally, in panel (d) we show the cycle-by-cycle prediction error.  Notice that this error is signed and defined as predicted minus observed maximum, {\it i.e.} a positive value corresponds to an overshot prediction.}
\label{fig:causalMaximumAmplitude}
\end{figure}

We now characterise the behaviour of the proposed methodology in a causal manner, {\it i.e.} by incremental incorporation of past solar cycles to the available data base.  Is the obtained linear relationship stable over time (in the sense of the stability of its coefficients) as new cycles become available?  We start from the earliest possible application of the proposed approach, namely predicting Cycle 3 with a linear response fitted to data from Cycles 1 and 2.  We then incorporate the actual data from Cycle 3 to the available data base and fit a linear model which we use to forecast the maximum amplitude of Cycle 4.  This exercise is repeated incrementally, always in a causal manner: to issue a prediction for a given solar cycle we only use information from previous cycles.  In this way we obtain the results presented in Figure \ref{fig:causalMaximumAmplitude}.  Panels (a) and (b) show the linear regression coefficients profile as a function of the predicted solar cycle number.  In panel (c) we plot the linear correlation coefficient and in (d) the signed prediction error (predicted minus observed maximum).  We observe that as new cycles are incorporated into the data base the proposed method quickly stabilises, and only small fluctuations are apparent after the tenth cycle. 

In \inlinecite{Verdes2000} we proposed the second-derivative prediction method and used it to issue a forecast for the maximum amplitude of Solar Cycle 23.  
The predicted maximum level was $115 \pm 32$, a value which was consistent with the climatological mean of historical data.  As mentioned in Section \ref{s:data}, in that work we used a more involved smoothing approach, while here we adopted a simple Savitzky--Golay scheme.  However, the different smoothing procedure employed in this work had no effect on the issued prediction.  Indeed, the forecast would be again 115, with a 90\% confidence interval of 94\,--\,143.  Cycle 23 finally showed a peak sunspot number of $119.6$, in remarkable agreement with the prediction, a coincidence which can be considered as exceptional rather than the rule.  Notice from Figure \ref{fig:linearFitMaximumAmplitude} that, together with Cycle 13, Solar Cycle 23 happened to be amongst the closest to the regression line -- for Cycle 1 no model was yet available.

\subsection{Time of Occurrence}
\label{ss:timeOfMaximumPrediction} 
\begin{figure} 
\centerline{\includegraphics[width=0.65\textwidth]{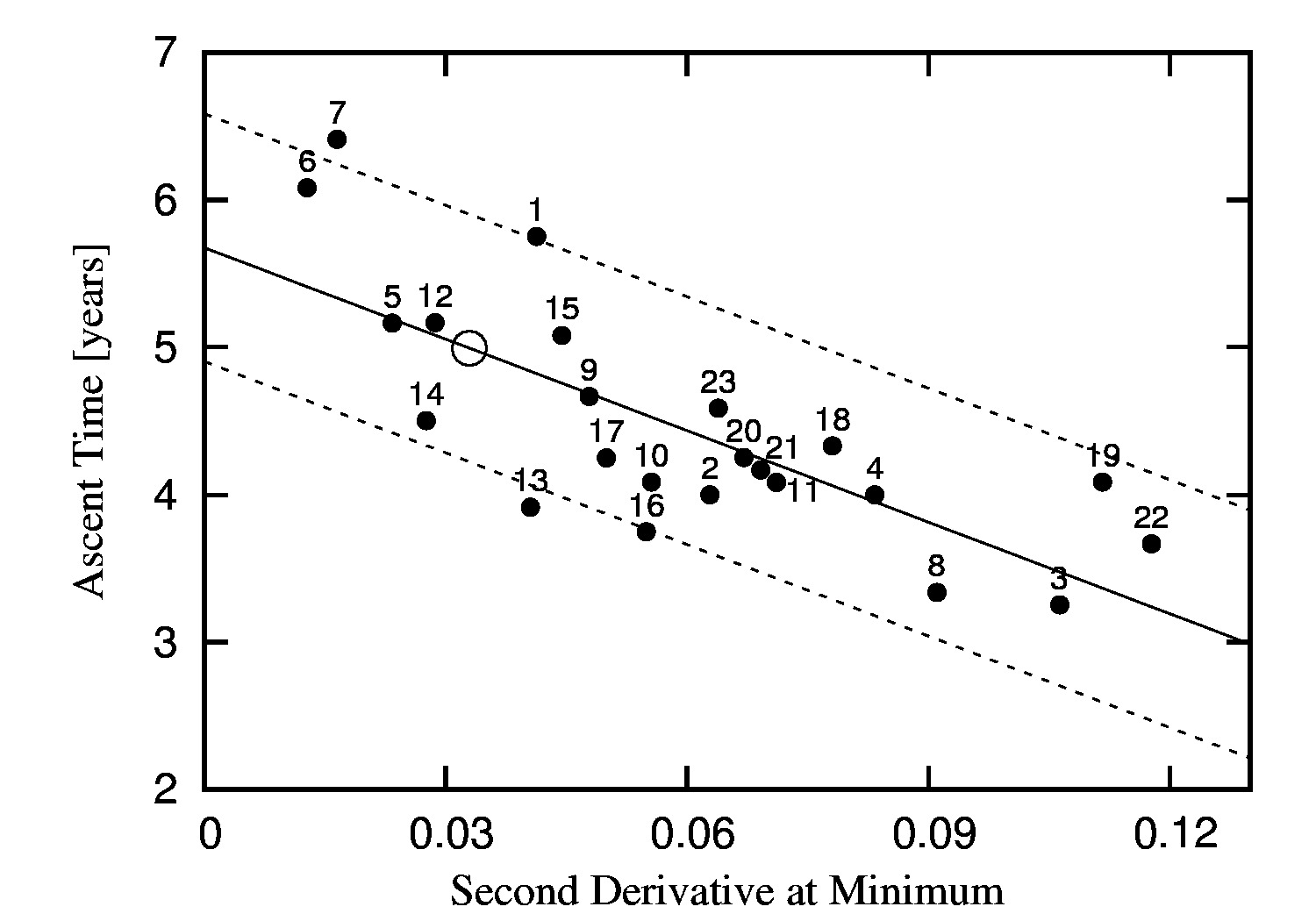}}
\caption{Ascent time of solar cycles against the value of the second derivative at the starting minimum.  As in Figure \ref{fig:linearFitMaximumAmplitude}, full and dashed lines are used to plot the result of a robust linear fit and its 90\% confidence interval, respectively.  Each dot is annotated by the solar cycle number it represents, and an open circle is used for the case of Solar Cycle 24.}
\label{fig:linearFitAscentTime}
\end{figure}

In order to predict when maximum solar activity will be observed, we measure the elapsed time between consecutive minima and maxima of $d_0(t)$.  Notice that we do not base the ascent time definition on the yearly averages time series, but prefer the finer resolution (monthly as opposed to yearly) of $d_0$ instead.  We regress the so-obtained ascent times to the curvature ($d_2$) at the corresponding minima.  In Figure \ref{fig:linearFitAscentTime} we show a scatter plot of these variables.  We find that the second derivative at the minimum has a good explanatory power of ascent time.  More precisely, they are inversely related, with a linear correlation coefficient of -0.77.  This inverse relationship is consistent with the rationale of the proposed approach: the larger $d_2$ at the minimum is, the faster the cycle rises and therefore the shorter the ascent time.

As Figure \ref{fig:linearFitAscentTime} shows, for Solar Cycle 24 we obtain a prediction of 5.0 years for the ascent time (shown with an open circle).  This interval is to be counted from the last minimum of our $d_0$ time series (October 2008), which brings the predicted sunspot number maximum of 78 to October 2013 (90\% confidence interval: January 2013 to September 2014).  The standard error of this prediction (sample standard deviation of residuals) is six months.

\begin{figure} 
\centerline{\includegraphics[width=0.75\textwidth]{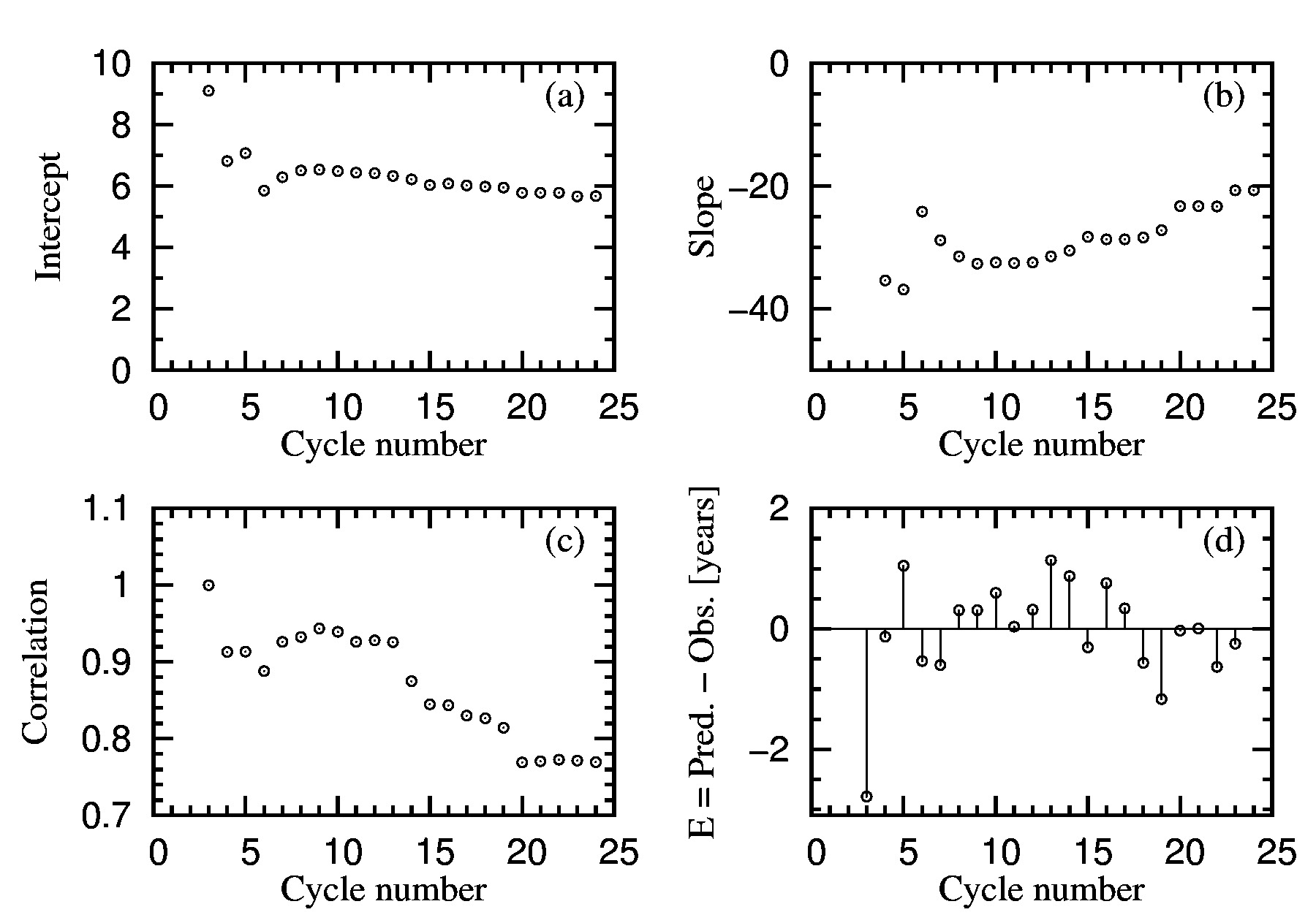}}
\caption{Characterisation of the proposed ascent time forecasting methodology.  In panel (a) we show the evolution of the linear regression-intercept as new cycles are added to the data base (more precisely, as a function of the predicted solar cycle number). Panel (b) profiles the slope behaviour. In (c) we plot the evolution of the correlation coefficient between maximum amplitude and curvature at the preceding minimum.  Finally, in panel (d) we show the cycle-by-cycle prediction error.  This error is signed and defined as predicted minus observed maximum, {\it i.e.} a positive value corresponds to an overshot prediction. Notice that the last points in panels (a), (b) and (c) correspond to the current coefficients and correlation of the model used to predict Solar Cycle 24.}
\label{fig:causalAscentTime}
\end{figure}

We characterise the proposed approach along the same lines employed in the previous section for the prediction of maximum cycle amplitude.  In Figure \ref{fig:causalAscentTime} we show how the unfolding of new cycles has affected the stability of the inverse relationship between ascent time and curvature at minimum reported here.  We find that the intercept has been broadly unaffected by new data.  In contrast, the slope was initially much steeper than it currently is.  As panel (b) of Figure \ref{fig:causalAscentTime} shows, Cycle 14 and more recently Cycles 19 and 22 have had a significant influence reducing the slope of the regression (the contribution of these cycles to the linear fit can be visualised in Figure \ref{fig:linearFitAscentTime}, where each dot is annotated by the solar-cycle number that it represents).

\subsection{Total Cycle Length}
\label{ss:lengthOfCyclePrediction} 

Finally, we use the same technique to address the problem of predicting solar cycle length, which we compute as the temporal interval between consecutive minima in $d_0(t)$.  As for the ascent time case, we prefer to measure temporal intervals on the monthly resolution of $d_0(t)$ instead of the coarser one available from the yearly averages time series.

\begin{figure} 
\centerline{\includegraphics[width=0.65\textwidth]{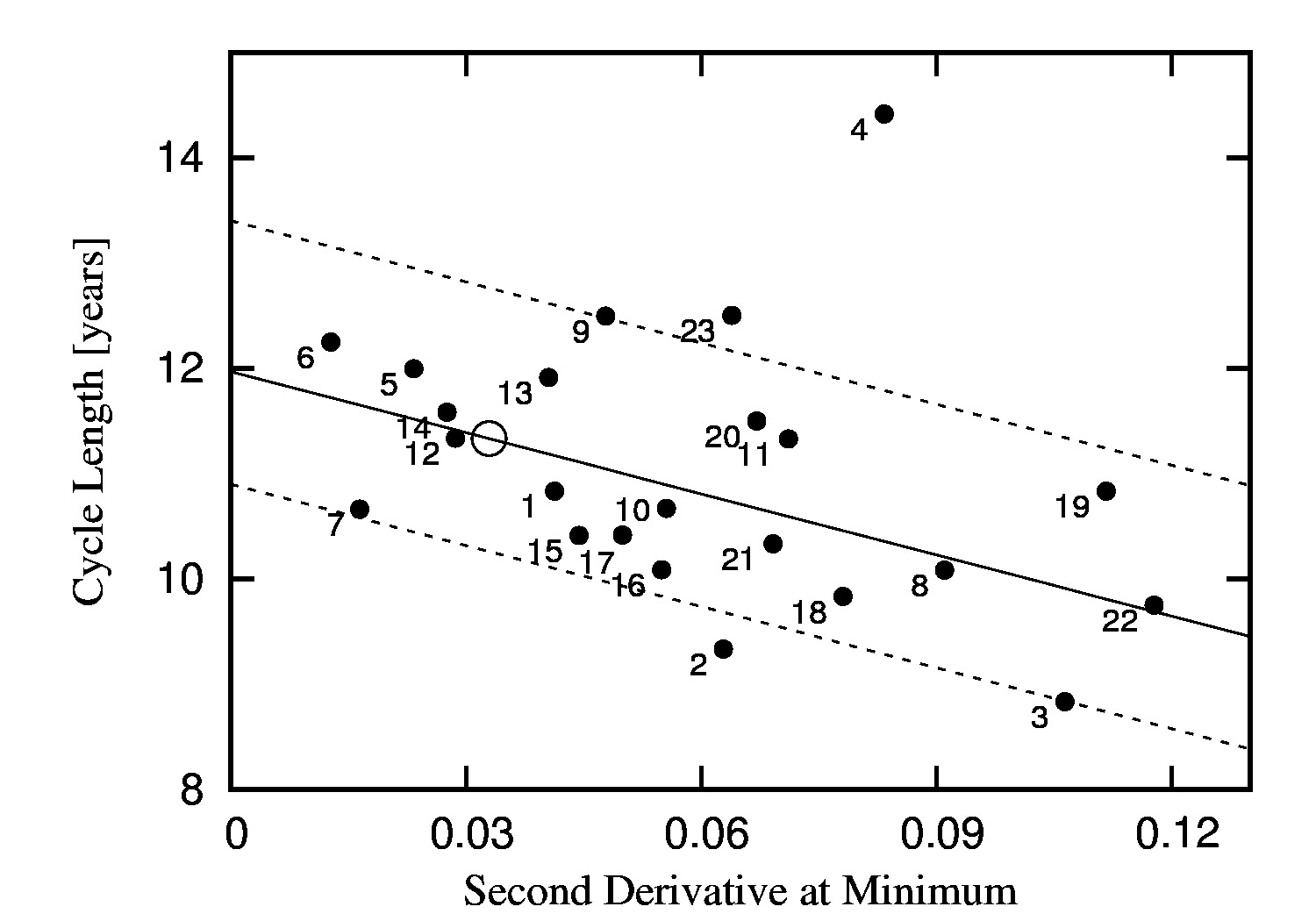}}
\caption{Total cycle length of solar cycles against the value of the second derivative at the leading minimum.  Full and dashed lines indicate a robust linear fit and the 90\% confidence interval, respectively.  Solar cycles are highlighted on the scatter plot by their number.  An open dot is used for the current cycle.}
\label{fig:linearFitCycleLength}
\end{figure}

We therefore regress total cycle length to curvature at minimum.  Inspection of these data, shown in Figure \ref{fig:linearFitCycleLength}, reveals the unusually long character of Cycle 4.  Indeed, there is some controversy in the literature about the nature of this cycle.  On one hand, it has been argued that the fourth solar cycle (1784\,--\,1798) is actually a superposition of two shorter cycles (\citeauthor{Usoskin2001}, 2001, 2003; \citeauthor{Usoskin2009}, 2009).  However, arguments against this assertion have also been elaborated \cite{Krivova2002,Zolotova2011}.  \inlinecite{Krivova2002} performed statistical tests of sunspot records, $^{10}$Be, $^{14}$C, and auroral proxy data to conclude that the hypothesis of two shorter cycles is not supported by existing data.  \inlinecite{Zolotova2011} reached the same conclusion and suggested that the length of the fourth cycle can be explained by impulse activity in the northern declining phase of the cycle.  In any case, since the validity of this data point to our analysis is under question, we opt to exclude it from the model building process.  Therefore, in the study of solar-cycle length {\it versus} curvature at minimum that follows below we do not include Solar Cycle 4.

Figure \ref{fig:linearFitCycleLength} reveals that the second derivative at the starting minimum carries information on the total length of the cycle.  As it was the case for ascent time, they are inversely related -- the linear correlation coefficient being in this case -0.56.  This inverse relationship is again consistent with the concept that larger curvatures at the leading minima are associated with stronger, shorter cycles.  In particular, for Solar Cycle 24 we obtain a prediction of 11.33 years for total length (shown with an open circle).  This interval is to be counted from the most recent minimum of the $d_0$ time series (October 2008), which brings the predicted end to February 2020 (90\% confidence interval: January 2019 to July 2021).  The standard error (sample standard deviation of residuals) is ten months.

\begin{figure} 
\centerline{\includegraphics[width=0.75\textwidth]{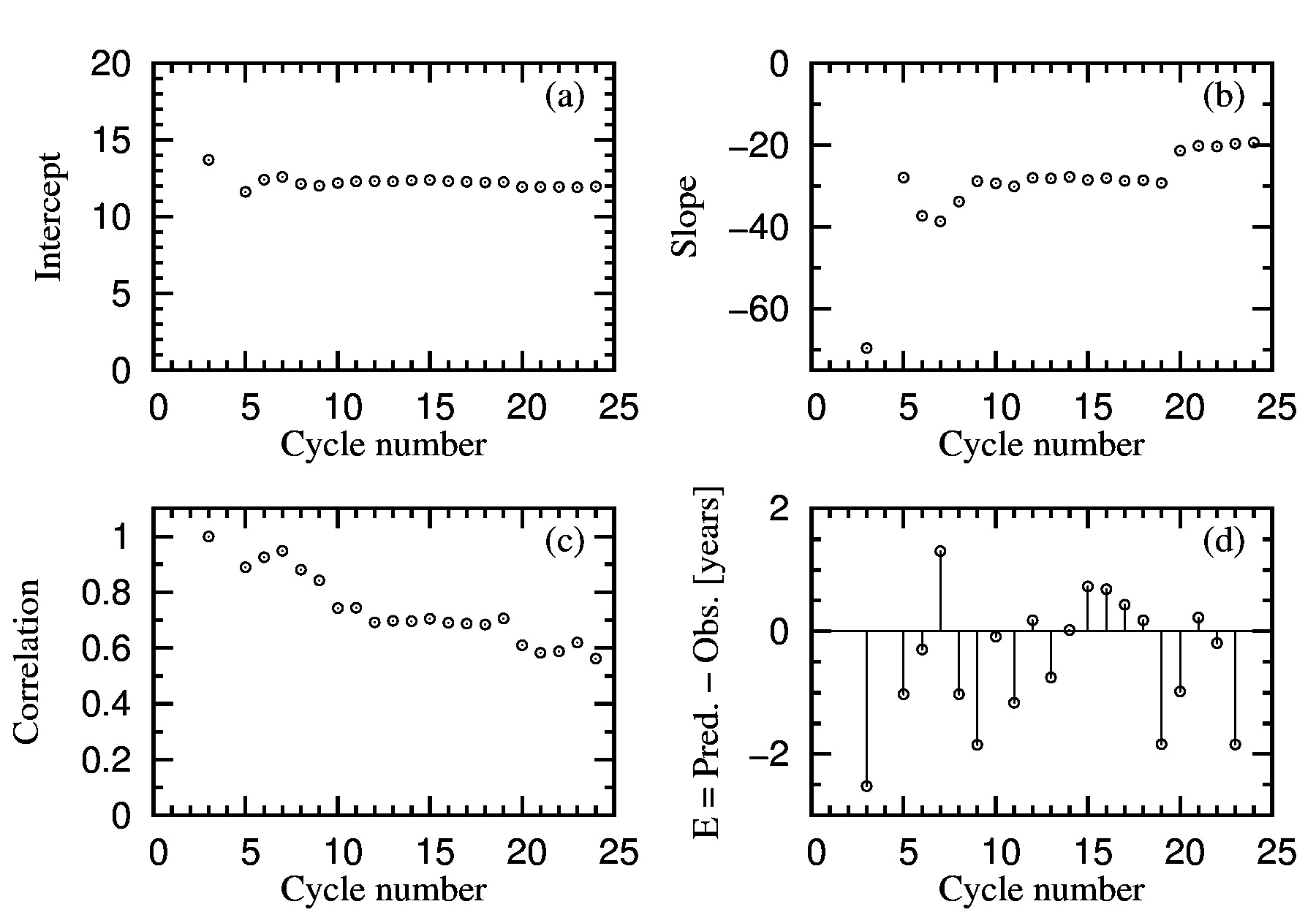}}
\caption{Unfolding of the proposed forecasting methodology in time.  Panel (a) profiles the evolution of the linear-regression intercept as new cycles are incorporated to the data base. Panel (b) shows the evolution of the slope. In (c) we depict the correlation coefficient between total cycle length and curvature at the starting minimum.  Finally, in panel (d) we plot the cycle-by-cycle prediction error.}
\label{fig:causalCycleLength}
\end{figure}

We complete the solar-cycle length study with the same incremental, causal analysis of its stability.  In Figure \ref{fig:causalCycleLength} we show how the unfolding of new cycles has progressively shaped the inverse relationship between total length and curvature at the leading minimum reported here.  We find that the intercept has been very consistent through time.  As panel (b) of Figure \ref{fig:causalCycleLength} shows, the extreme character of Cycle 19 had a significant influence redefining the slope of the regression, which has subsequently remained stable.

\subsection{Summary}
\label{ss:summary} 

\begin{figure} 
\centerline{\includegraphics[width=\textwidth]{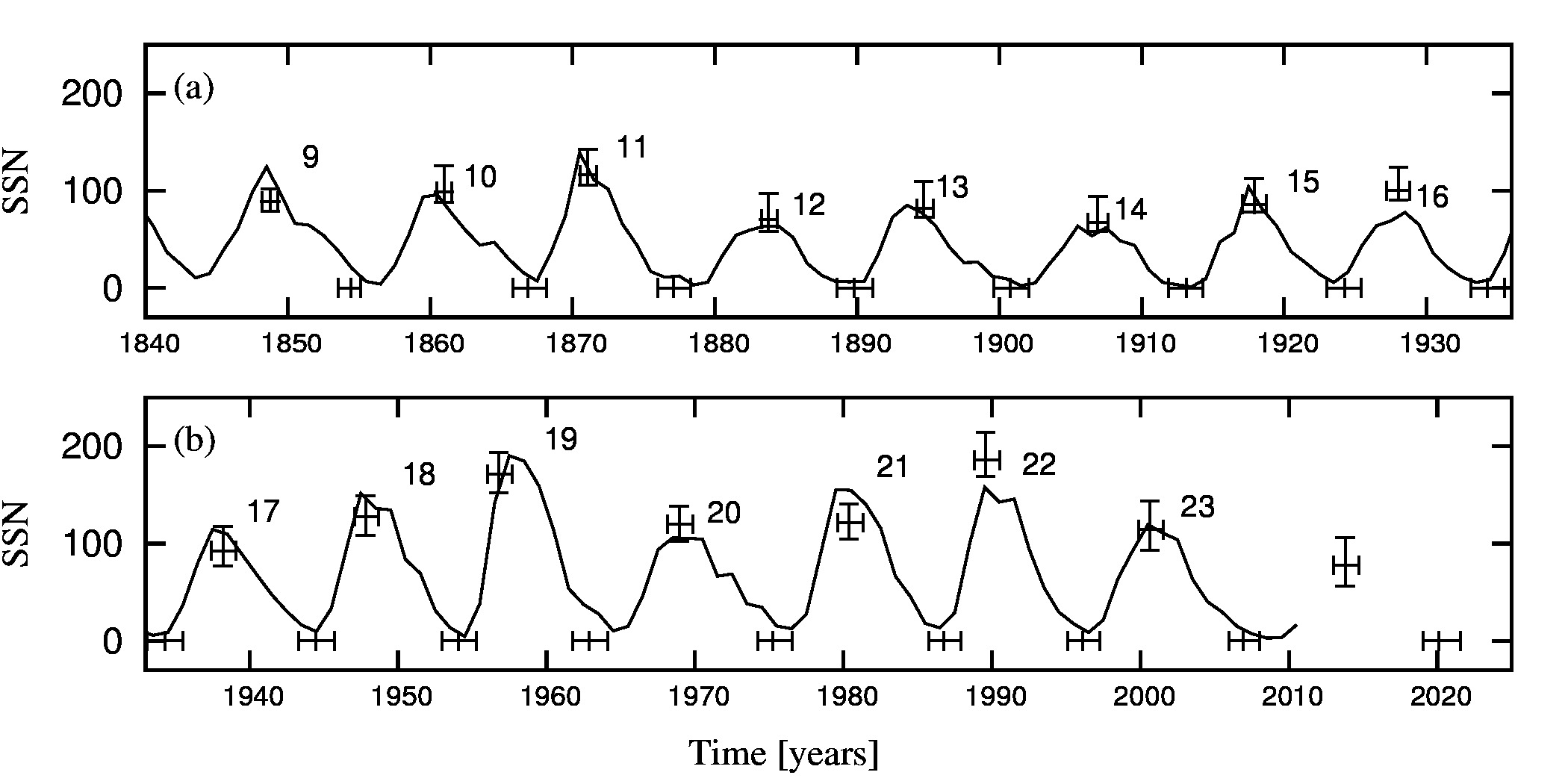}}
\caption{Predictions issued with the proposed methodology for the different cycles compared to the yearly average SSN time series (solid line).  Forecasts of maximum amplitude, timing and length (and their associated 90\% confidence intervals) are shown with bars and were obtained following a sequential protocol (see the main text for details).}
\label{fig:causalPredictions}
\end{figure}

In this subsection we collect the predictions obtained above into a consolidating view of forecasted solar-cycle features through time.  In Figure \ref{fig:causalPredictions} we plot the time series of yearly averages of sunspot numbers together with predictions elaborated with the methodology presented in this work.  Forecasts have been made in a purely causal manner by always using information from the past.  Every prediction (and its 90\% confidence interval) is drawn from a regression to curvature at the starting minimum that only sees the available data from previous cycles.  Notice the expanding character of confidence intervals as new cycles are incorporated to the proposed linear models.  
In the case of solar-cycle length predictions we use horizontal bars on the time axis (with an ordinate of 0).  The most recent bars depict our predictions for Solar Cycle 24.

\section{Conclusions}

In this work we have applied a previously developed methodology for maximum solar activity prediction \cite{Verdes2000} to the case of the current Solar Cycle 24, forecasting a cycle of moderate intensity (maximum sunspot number of 78).  Furthermore, we have generalised it to study the problem of maximum and subsequent minimum intensity timing.  The forecast of the latter features of solar activity bears a particular relevance, amongst others, for space mission planning, and space weather and climate change analyses.

The rationale behind the proposed methodology can be summarised in the following way: a fast-rising cycle (as measured by the curvature or second derivative at its starting minimum) is likely to be more intense than a slower-rising one.  By the same intuition, its ascent time and total cycle length will be shorter.  We have indeed found that the curvature at the leading minimum determines the main characteristics of the unfolding cycle.  We have also demonstrated the stability through time of the proposed predictive models.  We believe that the approach to solar-activity forecasting presented in this work constitutes a valuable addition to the existing body of statistical prediction methods in the literature.

%
\begin{acks}
We would like to thank an anonymous reviewer for his useful suggestions to improve this manuscript.  Financial support by CONICET, ANPCYT/MINCYT and Universidad Nacional de Rosario is gratefully acknowledged.
\end{acks}

\bibliographystyle{spr-mp-sola}
\bibliography{uzal}  

\end{article} 
\end{document}